\documentclass[twocolumn,superscriptaddress,floatfix,preprintnumbers,prl ,nofootinbib]{revtex4}
\usepackage[colorlinks=true,breaklinks=true]{hyperref}

\usepackage[utf8]{inputenc}
\hypersetup{allcolors=[rgb]{0.0 0.0 0.6},linkcolor=[rgb]{0.75 0.05 0.05}}
\usepackage{mathtools}
\usepackage[dvipsnames]{xcolor}
\usepackage{float}
\usepackage{siunitx}
\usepackage{letltxmacro}
\LetLtxMacro{\oldcite}{\cite}
\renewcommand{\cite}[1]{\mbox{\oldcite{#1}}}

\DeclareSIUnit\electronvolt{e\kern-.05em V}
\DeclareSIUnit\parsec{\text{pc}}
\DeclareSIUnit\clight{\text{\ensuremath{c}}}

\usepackage{orcidlink}

\long\def\exclude#1{}

\def\gammaHI{\Gamma_{\rm HI}}

\def\xe{x_{\rm e}}

\newcommand{\beq}{\begin{equation}}
\newcommand{\eeq}{\end{equation}}

\def\ga{\,\,\raise0.14em\hbox{$>$}\kern-0.76em\lower0.28em\hbox
{$\sim$}\,\,}

\newcommand{\taureio}{\tau_{\mathrm{e}}}

\allowdisplaybreaks

\bibpunct{[}{]}{,}{n}{}{,}

\usepackage[colorinlistoftodos]{todonotes}

\begin{document}

\title{Constraints on the Thomson optical depth to the CMB from the Lyman-$\alpha$ forest}

\author{Olga Garcia-Gallego\,\orcidlink{0009-0009-9003-7889}}
\email{og313@cam.ac.uk}
\affiliation{Institute of Astronomy, University of Cambridge, Madingley Road, Cambridge CB3 0HA, UK}
\affiliation{KICC - Kavli Institute for Cosmology Cambridge, Madingley Road, CB3 0HA Cambridge, United Kingdom}
\author{Vid Ir\v{s}i\v{c}\,\orcidlink{0000-0002-5445-461X}} \email{v.irsic@herts.ac.uk}
\affiliation{KICC - Kavli Institute for Cosmology Cambridge, Madingley Road, CB3 0HA Cambridge, United Kingdom}
\affiliation{Center for Astrophysics Research, Department of Physics, Astronomy and Mathematics, University of Hertfordshire, College Lane, Hatfield AL10 9AB, UK}

\author{Martin G. Haehnelt\,\orcidlink{0000-0001-8443-2393}}
\affiliation{Institute of Astronomy, University of Cambridge, Madingley Road, Cambridge CB3 0HA, UK}
\affiliation{KICC - Kavli Institute for Cosmology Cambridge, Madingley Road, CB3 0HA Cambridge, United Kingdom}

\author{James S. Bolton\,\orcidlink{0000-0003-2764-8248}}
\affiliation{School of Physics and Astronomy, University of Nottingham, University Park, Nottingham, NG7 2RD, UK}


\begin{abstract}
We present the first constraints on the electron optical depth to reionization, $\taureio$, from the Lyman-$\alpha$ forest alone for physically motivated reionization models that match the reionization's end-point, $z_{\rm{end}}$, required by the same astrophysical probe, and for symmetric reionization models with fixed duration, $\Delta z$, commonly adopted in CMB reionization analyses. Compared to traditional estimates from the latter, the Lyman-$\alpha$ forest traces the ionization state of the IGM through its coupling with the thermal state. We find an explicit mapping between the two solving the chemistry and temperature evolution equations for hydrogen and helium. Our results yield $\taureio$=$0.040^{+0.042}_{-0.018}$ (95\% C.L) and $\taureio$=$0.041^{+0.028}_{-0.017}$ for reionization models with $z_{\rm{end}}$ and $\Delta z$-fixed, respectively. With mock Lyman-$\alpha$ forest data that mimics the precision of future larger quasar sample datasets, we would potentially obtain tighter $\taureio$ constraints, paving the way for CMB-independent constraints on the epoch of reionization from a large-scale structure probe.


\end{abstract}

\maketitle

\textit{Introduction.}--- The optical depth to reionization, $\taureio$,  has recently drawn considerable attention due to its potential to significantly weaken the hints for physics beyond $\Lambda$CDM suggested by the latest results from the Dark Energy Spectroscopic Instrument (DESI) Data Release 2. When combined with Planck Cosmic Microwave Background (CMB) measurements and supernovae datasets, baryon acosutic oscillations (BAO) from DESI DR2 reveals a $\sim$ 4$\sigma$ preference for dynamical dark energy (\cite{desiDR2}). Primarily driven by a lower value of $\Omega_{m}$, this in turn implies a preference for neutrino masses below the minimum allowed by neutrino oscillation experiments (\cite{elbers25}). These findings have motivated recent work by \cite{sailer25, jhaveri25} to consider the implications of larger optical depths than the standard estimate obtained from polarization measurements at low$-l$ multipoles of the CMB, which yields $\taureio$ = 0.054$\pm$0.007 (68\% C.L.) (\cite{Planck18}). Since the CMB polarization signal is roughly 100 times weaker than the corresponding temperature fluctuations and subject to the same levels of foreground contamination, $\taureio$ is the least well-constrained parameter within the $\Lambda$CDM model. In fact, \cite{sailer25, jhaveri25} found that a value of $\taureio$ $\approx$ 0.09 restores consistency between Planck and DESI,  reducing the statistical tension below $\sim$ 2$\sigma$. Notably, such \textit{high} $\taureio$ values are consistent with early WMAP results \cite{bennet13}, and with recent constraints from the combination of small-scale CMB anisotropies and CMB lensing with various low-redshift probes (\cite{elbersnew}). 

It is important to highlight that, unlike cosmological parameters such as $\Omega_{m}$, $\taureio$ is fundamentally astrophysical, encoding information about the integrated ionized electron fraction up to the epoch of decoupling. To interpret different $\taureio$ estimates, it is crucial to revisit our current understanding of the Epoch of Reionization (EoR), mainly constrained through the Lyman-$\alpha$ forest. Observational evidence, such as the presence of Gunn-Peterson troughs (\cite{fan06, mcgreer15}), large-scale opacity variations (\cite{bosman22}), and Lyman-$\alpha$ damping wings (\cite{becker24}), at $z$ $<$ 6, along with short mean free paths below this redshift (e.g. \cite{becker21}), places robust constraints on a \textit{late} end-point of reionization, $z_{\rm{end}}$. At the same time, recent JWST detections of Lyman-$\alpha$ emitters (LAEs) at high-$z$ appear to suggest an \textit{early} onset of reionization (e.g. \cite{witsok25}), indicating on-going reionization already by $z$=13, as discussed by \cite{cohon25}. 
\begin{figure*}[hbtp!]
    \centering
    \includegraphics[width=\linewidth]{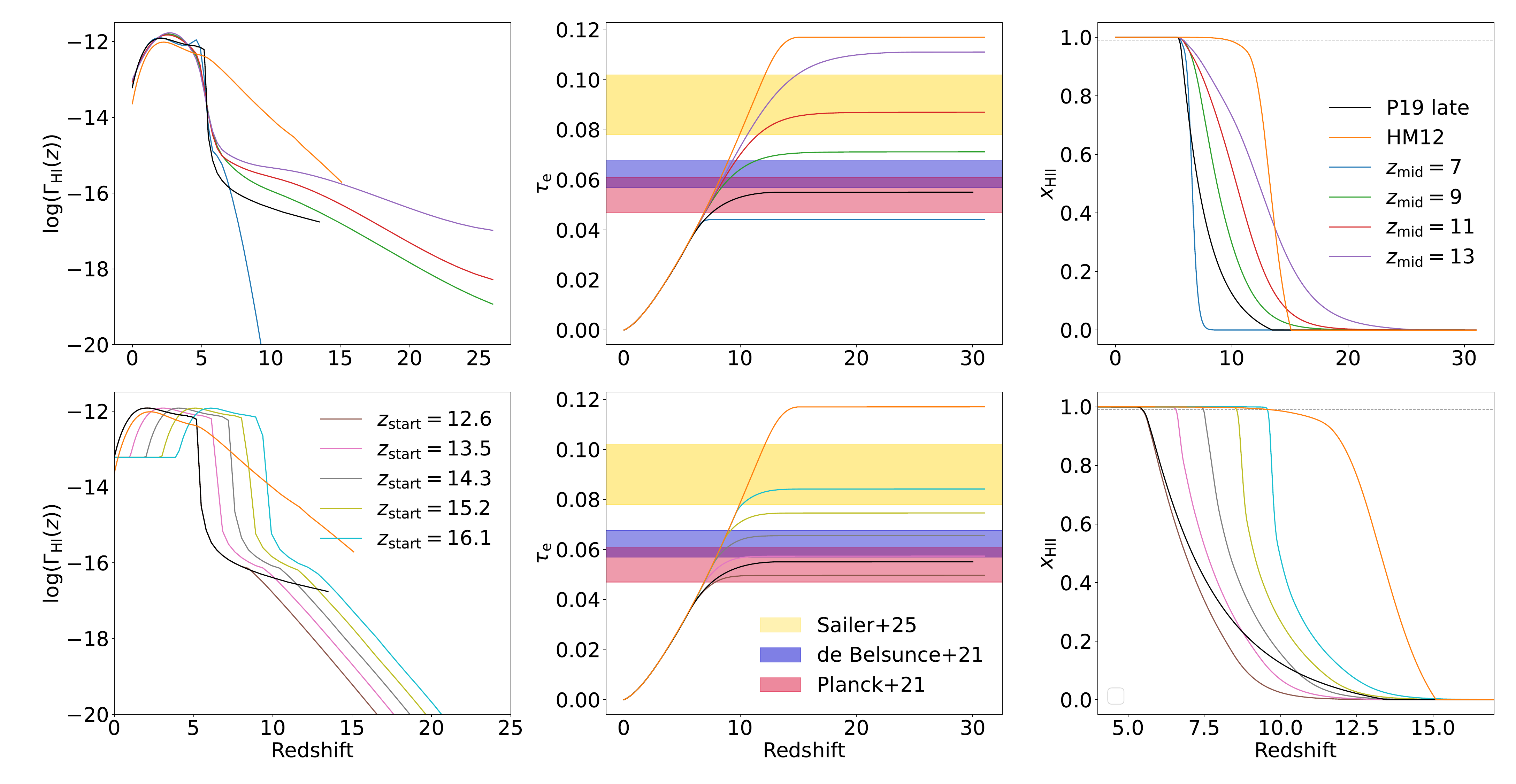}
    \caption{H I photo-ionization rate $\gammaHI$ (left), electron scattering optical depth $\taureio$ (middle) and H II ionized fraction $x_{\rm{HII}}$ (right) for reionization models that keep $z_{\rm{end}}$=${z}_{\rm{end}}^{\rm{ P19\,late}}$ fixed (varying $z_{\rm{mid}}$) and $\Delta z$ $\approx$ 5.5 fixed (varying $z_{\rm{start}}$) in the top and bottom rows, respectively. We highlight in the middle column $\taureio$ inferred by previous works (\cite{Planck18, belsunce21, sailer25}). We further show \cite{puchwein19}'s late and \cite{hm12}'s reionization models in black and orange, respectively. The gray dashed line on the right column shows the Lyman-$\alpha$ forest end-point of reionization requirement from the former: $x_{\rm{HII}}$ $\approx$ 1. } \label{grid}
\end{figure*}
Taking the Lyman-$\alpha$  constraints for the late reionization stages into account, \cite{cain25} used measurements of the patchy kinetic Sunyaev Zel’dovich (pkSZ) effect to place limits on the duration of reionization. Their results indicate that reconciling Lyman-$\alpha$ measurements with a \textit{high} $\taureio$ leads to a $\geq$ 2$\sigma$ tension. However, the pkSZ is highly sensitive to foreground subtraction (\cite{reichard21}) and the measurement depends on the galaxy clustering model assumed (\cite{garrett25}). Such \textit{high} values of $\taureio$ have been excluded with even more significance by \cite{elbersnew} using a compilation of astrophysical measurements of $x_{\rm{HII}}$ from damping wings and dark pixel constraints. These allow to reconstruct the reionization history without assuming a particular parametric model, leading to $\taureio$ constraints that are independent of CMB data.

In this \textit{Letter}, we argue that current post-reionization Lyman-$\alpha$ forest data can independently determine the allowed $\taureio$ for a range of physically motivated reionization models, and for symmetric reionization models, extensively studied by CMB analyses.
The constraining power comes from the connection between $\taureio$ and the thermal history of the IGM, to which the Lyman-$\alpha$ forest is highly sensitive: during reionization, baryons react to the photo-heating of the IGM and the corresponding increase of the gas pressure, suppressing small-scale structure in the Lyman-$\alpha$ forest. Assuming a cold dark matter cosmology, \cite{boera19} provided the first constraints on these two effects using hydrodynamical simulations for a wide range of thermal history models. The results, together with those from subsequent work (\cite{irsic23, gg25}), agree remarkably well with the observational evidence for a $late$ end to reionization and independent measurements of the thermal evolution of the IGM (\cite{gaikwad20, gaikwad21}). Using an empirical mapping between the Lyman-$\alpha$ forest thermal parameters and $\taureio$ from our simulations, we constrain the history of the EoR through $\taureio$, independently of CMB observations. 
\begin{figure*}[hbtp!]
    \centering
    \includegraphics[width=0.35\linewidth]{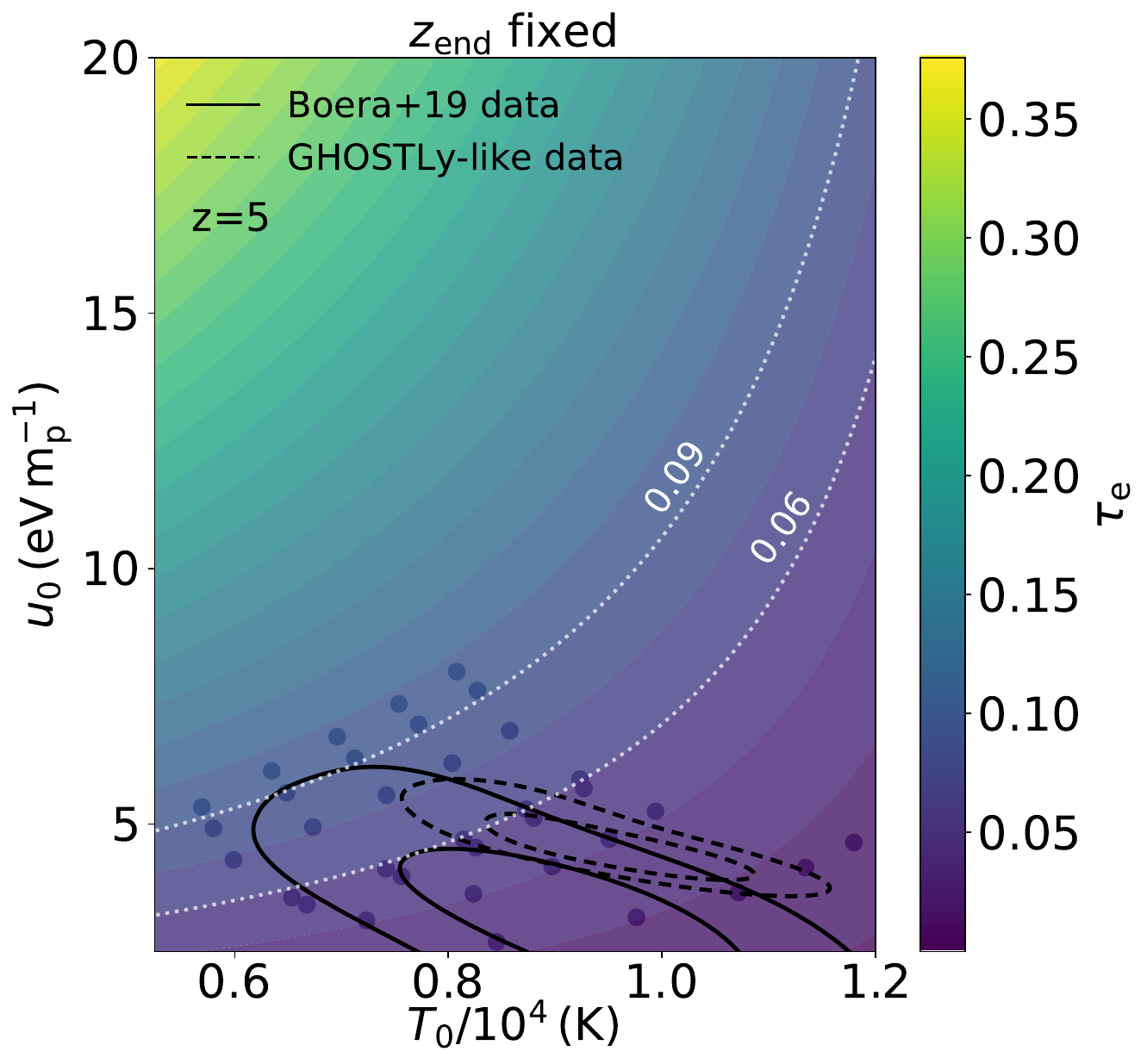}
    \includegraphics[width=0.35\linewidth]{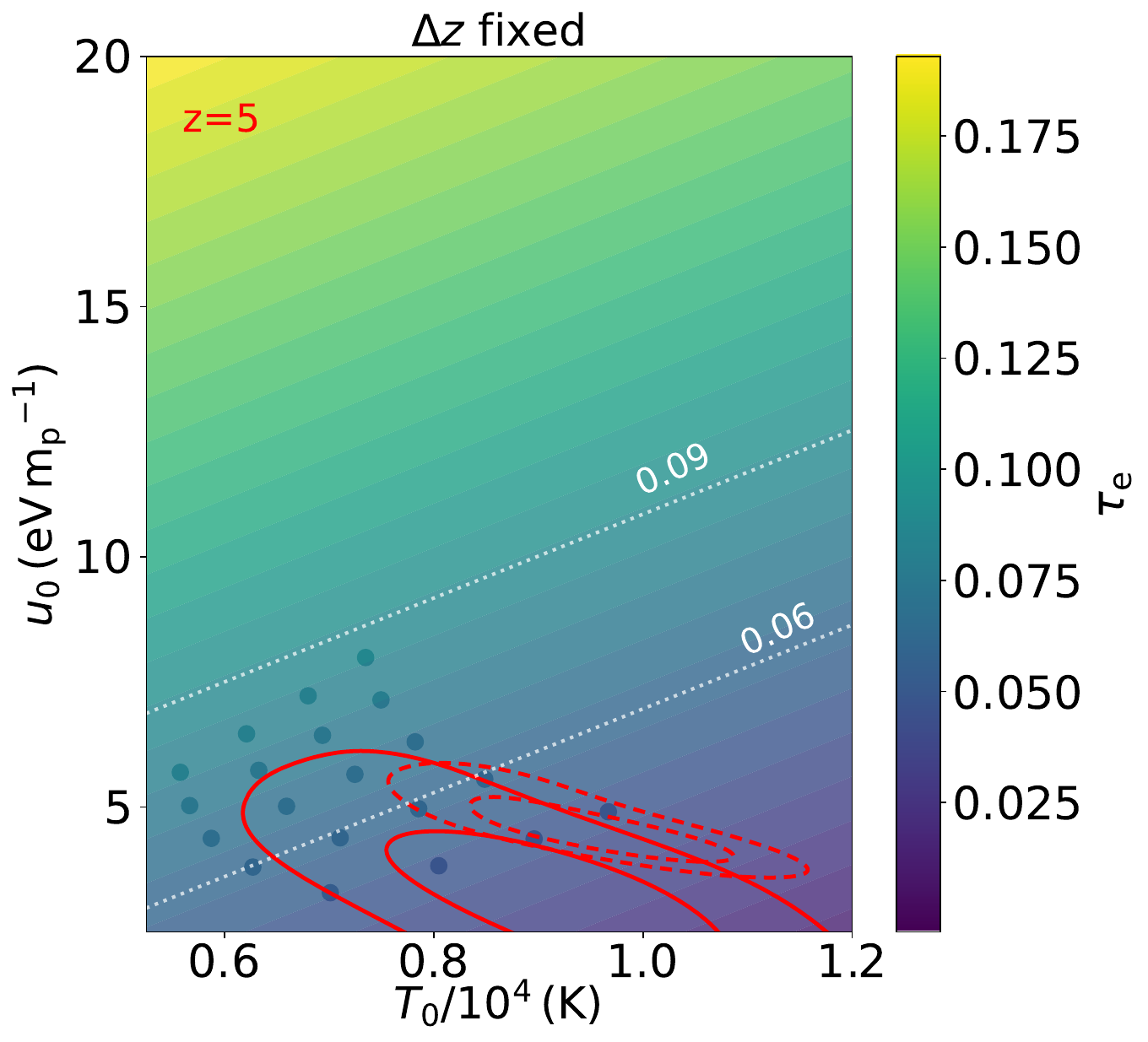}
    \caption{Posterior in the $u_{0}-T_{0}$ plane at $z$=5.0 for the default analysis in \cite{gg25} (Gaussian $T_{0}$ priors) shown as black and red solid contours. The posteriors resulting from the chains using mock data with smaller relative error bars of 5\% are shown as dashed contours also in black and red. The colormap shows isocontours for $\taureio$. We highlight the $\taureio$ = [0.06, 0.09] contours in white. The grid points show the simulated reionization models used to infer the extrapolation scheme for $\taureio$ with: fixing $z_{\rm{end}}$ in the \textit{left} and $\Delta z$ in the \textit{right}. The grid points along the $\taureio$ isocontours are obtained by rescaling the original photo-heating rates by $\alpha$ = [0.5,0.6,0.7,0.8,0.9] \cite{villasenor22}.} \label{isocontours}
\end{figure*}

\textit{Thermal and ionization state of the IGM}--- 
The ionization state of the IGM is in this work constrained through the electron opacity to CMB photons, $\taureio$, which depends on the line-of-sight integral of the electron fraction, $\xe$, with its main contribution coming from the redshift range of the EoR. On the other hand, the IGM's thermal history is described by the cumulative energy deposited into the gas per proton by photo-heating, $u_0$ (\cite{nasir16}). This quantity is an integral of the total photo-heating rate per unit volume, $\mathcal{H}_{i}=\sum{n_{i}\epsilon_{i}}$, where $n_{i}$ and $\epsilon_{i}$ are the number densities and photo-heating rates, respectively, for the species $i$ $\in$ [H I, He I, He II]. The $u_0$ integration redshift range is chosen here to match the time scales at which the flux power spectrum is more sensitive to heating (see Section 6.3.3 in \cite{boera19}). Note that $u_0$ is only weakly correlated with thermal broadening, parameterized by the gas temperature at mean density, $T_0$. The connection between the thermal parameters $u_0-T_0$ and $\taureio$ comes from the relation,
\begin{equation}
\mathcal{H} = n_{\rm{HI}}\epsilon_{\rm{HI}}=n_{\rm{HI}}E_{\rm{ion}}\gammaHI, \label{Eion}
\end{equation}
where $\gammaHI$ and $\epsilon_{\rm{HI}}$ are the hydrogen photo-ionization and photo-heating rates, respectively, and $E_{\rm{ion}}$ corresponds to the mean excess energy available to photoheat the IGM (\cite{puchwein19}), typically assumed to be constant (\cite{mcquinn15}). The Lyman-$\alpha$ forest, therefore, probes $\gammaHI$ through its coupling with $\epsilon_{\rm{HI}}$. At the same time, the photo-ionization rate, $\gammaHI$, enters into the definition of $\taureio$ through the evolution of
$\dot{\xe}$, or similarly, of the hydrogen ionized fraction, $\dot{x_{\rm{HII}}}$. The latter is given by the well-known “reionization equation” (\cite{madau99}), which, written in terms of the fractional abundances (Eq.(2) in \cite{madau17}) shows the connection
between $\gammaHI$ and $\xe$, and therefore, $\taureio$. Hence, Eq.~(\ref{Eion}) implies that the Lyman-$\alpha$ forest contains information on $\taureio$ and that the parameter pair $u_0-T_0$ can be mapped to $\taureio$. 

\textit{Data}--- 
We use the 1D Lyman-$\alpha$ flux power spectra dataset presented in \cite{boera19}. The measurements are from a high-resolution sample of 15 quasars observed by HIRES and UVES at $z_{\rm{bin}}$=[4.2, 4.6, 5.0]. For more details on the data, we refer the reader to \cite{irsic23, gg25}. Given the moderate number of quasar sightlines in this sample, we further consider a second mock dataset constructed from synthetic Lyman-$\alpha$ forest spectra extracted from Sherwood-Relics simulations' skewers. The mock covariance matrix is rescaled to mimic 5\% relative uncertainties. This mock dataset allows us to forecast how sensitive future 1D flux power spectra extracted from a large sample of quasars at the same redshift range (4$<$$z$$<$5) would be to $\taureio$. The GHOSTLy (Gemini High Resolution Optical Spectrograph) survey is a program that will target $\sim$30 quasars, achieving comparable flux power spectrum error bars to that from our mock data \cite{artola24, kalari24}. Hereafter, analyses using the mock dataset will be termed GHOSTLy-like.   

\textit{Hydrodynamic Simulations/RT code.}---The non-linear physical processes of the IGM to which the Lyman-$\alpha$ forest is sensitive can only be  interpreted quantitatively using hydrodynamical simulations. In our previous work (\cite{gg25}), we have used the Sherwood-Relics suite of simulations (\cite{puchwein23}) with varying astrophysical parameters, including the mean IGM transmission $\tau_{\rm{eff}}$ and temperature-density power-law index $\gamma$ along with $T_0$ and $u_0$, to train a neural network emulator. Cosmological parameters (e.g. $\sigma_8$) are kept fixed, to which the astrophysical parameters $u_0, T_0$ are insensitive \cite{irsic17}. We build on previous  work by considering the late reionization model from \cite{puchwein19}, with $z_{\rm{end}}^{\rm{ P19\,late}}$= $z$($x_{\rm{HII}}$ $\approx$ 1).
We further consider redshift-symmetric reionization models with fixed duration $\Delta z$, corresponding to the usual EoR parameterization employed in CMB analyses. To explore reionization models matching the constraints on $z_{\rm{end}}$ and $\Delta z$ separately, we use the non-equilibrium photoionization code developed by \cite{bolton22}, and recently used by \cite{soltinsky21}. Given input photo-ionization rates $\Gamma_{i}$ and photo-heating rates $\epsilon_{i}$, the code solves the four coupled first order differential equations for the abundance of ionized hydrogen and helium (the full “reionization equation" for each species), and for the temperature evolution (see Eqs.(B5)-(B8) in \cite{bolton07}). We determine $\Gamma_{\rm{HI}}$ and $\epsilon_{\rm{HI}}$ by trial-and-error, solving the non-equilibrium code until the resulting $\xe$ evolution matches reionization models with $z_{\rm{end}}$ = ${z}_{\rm{end}}^{\rm{P19\,late}}$-fixed and $\Delta z$-fixed (see tabulated $\gammaHI(z)$ in Tables~I, II and corresponding $\xe(z)$ in Figure S2 in the \textit{Supplemental Material} \cite{supplemental}). In practice, we use as a starting point a simplified version of Eq.(B5) in \cite{bolton07}, which relates $\gammaHI$ and $\xe$. The photo-heating rates are obtained by exploiting their coupling with $\Gamma_{\rm{HI}}$ from Eq.~(\ref{Eion}), with $E_{\rm{ion}} = {E_{\rm{ion}}}^{\rm{P19\,late}}$. The relevant He I quantities are effectively coupled to the H I rates, since both H I and He I reionization are driven by the same sources. The He II rates are left unchanged with respect to the late reionization model of \cite{puchwein19}, ensuring that the low-redshift evolution remains consistent with observational constraints.

\begin{figure*}[hbtp!]
    \centering
    \includegraphics[width=0.5\linewidth]{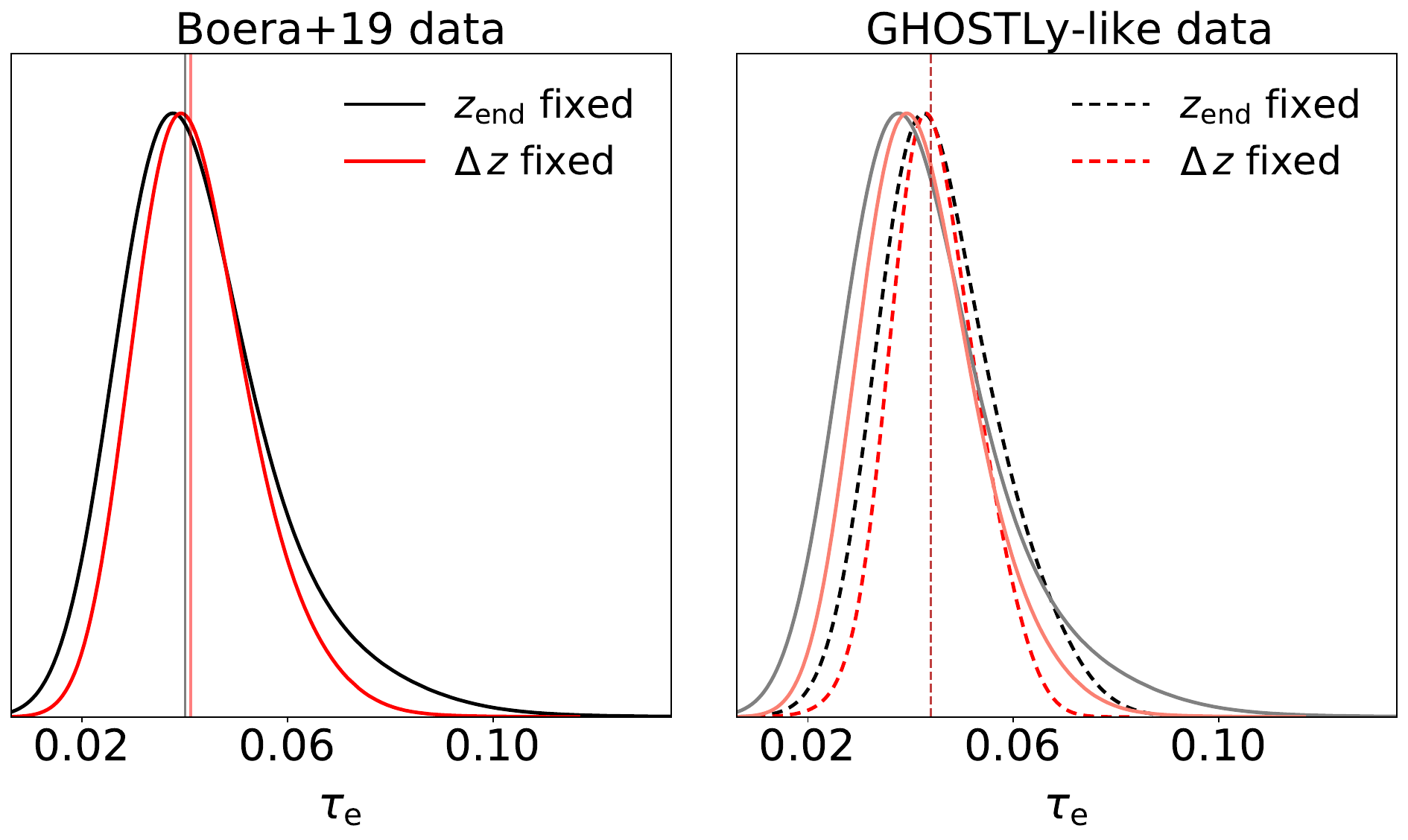}
    \caption{1D posterior distribution for $\taureio$ obtained with $z_{\rm{end}}$-fixed in black and $\Delta z$-fixed in red reionization models. Solid and dashed lines correspond to the analysis that uses \cite{boera19} and GHOSTLy-like data, respectively. Vertical lines indicate the median of each distribution.} \label{1d}
\end{figure*}

Our code calculates $x_{\rm{HII}}$, $x_{\rm{HeII}}$ and $x_{\rm{HeIII}}$, from which we compute $u_{0}$. We further integrate over $\xe$, tied to the evolution of the number density of  species $i$ (see the closing conditions in Appendix B3 in \cite{bolton07}), to compute $\taureio$. Finally, we extract $T_0$ at the redshift $z_{\rm{bin}}$ of the data. In Figure~\ref{grid}, we show the redshift evolution of $\gammaHI$ for these two sets of reionization models, and $\taureio$ and $x_{\rm{HII}}$ calculated from the output of the photo-ionization code as described above. 

To map the thermal and ionization parameters, (i.e. $u_0-T_0$ and $\taureio$), we first rescale $E_{\rm{ion}}$ by a factor $\alpha_{\rm{E}}$. This allows us to increase the number of simulated points used in the $u_0(z_{\rm{bin}})-T_0(z_{\rm{bin}})$ plane, to improve the precision of our inference. This transforms $\mathcal{H}_{\rm{HI}}$ following \cite{villasenor22} and leaves $\taureio$ unchanged. In this way, we obtain isocontours of $\taureio$ in the $u_0(z_{\rm{bin}})-T_0(z_{\rm{bin}})$ plane within the grid points. We further find that $z_{\rm{bin}}$ = 5.0 is the redshift from the \cite{boera19} data most sensitive to $\taureio$. From the definition of $\taureio$, $\xe\sim$ $\dot{\taureio}$, which depends on $\gammaHI$. From Figure~\ref{grid}, $\dot{\taureio}$ is largest at the end of reionization, and therefore the contribution to the cumulative heat peaks at the redshift close to this point (from Eq. (\ref{Eion}) one can see that the rate of change of $u_0$ depends on photo-heating rates, and therefore on $\xe$).

Therefore, we map $u_0-T_0$ at $z$ = 5.0 (${u_0}^{5.0}$, ${T_0}^{5.0}$) and $\taureio$ by fitting a linear relation using the grid points in Figure~\ref{isocontours} obtained from the photo-ionization code. Since reionization models with $z_{\rm{end}}$-fixed exhibit a tighter correlation with $u_{0}$ compared to the $\Delta z$-fixed models, we include a second-order term in the fit for the former, leading to the isocontours shown in Figure~\ref{isocontours}. 

\textit{Results.---} The main results are shown in Figure~\ref{isocontours} and Figure \ref{1d}. The former shows the relation in the $u_0^{5.0}-T_0^{5.0}$ plane between thermal parameters and $\taureio$. We further show the contours obtained for a CDM cosmology using the emulator from \cite{gg25} to check the range of $\taureio$ allowed by the \cite{boera19} data. Using the fit for $\taureio$ (${u_0}^{5.0}$, ${T_0}^{5.0}$) for each set of reionization models yields the following 68\% (and 95\% C.L.) constraints:  
\[
\left.
\begin{array}{ll}

{\taureio}=0.040{^{+0.017}_{-0.010}}\left.^{(+0.042)}_{(-0.018)}\right.\,\text{\,\,\,$z_{\rm{end}}$-fixed} \\
{\taureio}=0.041{^{+0.012}_{-0.010}}{^{(+0.028)}_{(-0.017)}}\,\text{\,\,\,$\Delta z$ fixed}
\end{array}
\right\}
\,\,\text{Boera+19.}
\]

From our mock data we furthermore find,
\[
\left.
\begin{array}{ll}
{\taureio}=0.044{^{+0.014}_{-0.010}}{^{(+0.027)}_{(-0.017)}}\,\text{\,\,$z_{\rm{end}}$-fixed} \\
{\taureio}=0.044{^{+0.010}_{-0.007}}{^{(+0.018)}_{(-0.014)}}\,\text{\,\,$\Delta z$ fixed}
\end{array}
\right\}
\,\,\text{GHOSTLy-like}.
\]

\textit{Discussion.---}
The results above indicate that independent constraints of $\tau_e$ from  Lyman-$\alpha$ forest data are in good agreement with the \textit{Planck} polarization measurements (\cite{Planck18}). They are also consistent with other work that has measured $\taureio$ without using CMB data \cite{elbersnew}, but note that our work does not rely on astrophysical constraints on $x_{\rm{HII}}$. \\
Compared to the $\taureio$ measurement from CMB and DESI BAO \cite{sailer25, jhaveri25}, we find a mild tension at the 2.4$\sigma$ and 2.8$\sigma$ level with current Lyman-$\alpha$ forest data for the $z_{\rm{end}}$ and $\Delta z$-fixed models, respectively. The tension weakens for $\taureio$=0.08 which, as suggested by \cite{sailer25}, would be sufficient to reconcile DESI with CMB observations.
The panel on the right of Figure~\ref{1d}  demonstrates the constraining power on $\taureio$ that will be achieved with forthcoming quasar samples. Importantly, the upper uncertainty on $\taureio$ will improve by a factor of $\sim$ 1.6. Consequently, if GHOSTLy-like data leads to similar constraints on the median of $\taureio$ as current data, a \textit{high} $\tau=0.09$ will be excluded with 2.5$\sigma$ and 3$\sigma$ significance, respectively, for the two reionization models considered.  Relating these results to the nature of the reionization history, we find that for \textit{late} reionization models (matching $z_{\rm{end}}$ inferred from the Lyman-$\alpha$ forest opacity), reionization also needs to be \textit{rapid}, since more extended histories would lead to a higher $\taureio$. For models with fixed duration $\Delta z$, \textit{early} reionization models are strongly ruled out by the flux power spectrum, since these will also imply a high $\taureio$. We assume a conservative $\Delta z$$\approx$ 5.5, compared to Planck \cite{Planck18}, since smaller values require earlier onset/end of reionization to achieve the same levels of $\taureio$, which is strongly disfavored by Lyman-$\alpha$ forest opacity constraints; Our baseline constraint, ${\taureio}=0.040^{+0.042}_{-0.018}$,  suggests that the Lyman-$\alpha$ forest flux power spectrum alone prefers a relatively \textit{rapid} and \textit{late} reionization history. However, the significance of these results depends on the  observational uncertainty on the value of $z_{\rm{end}}$, as well as on the assumed duration of reionization.
Such \textit{high} $\taureio$ values seemed to only be preferred due to the presence of strong degeneracies between the cosmological parameters, mainly with $\Omega_{\rm{m}}$ and $A_{s}$ {\cite{elbers25}}. Thus, we find that, in agreement with \cite{elbersnew,cain25,garrett25}, a large CMB optical depth cannot meet the Lyman-$\alpha$ forest constraints on $z_{\rm{end}}$.

We also note from Figure \ref{1d} that the Lyman-$\alpha$ flux power spectrum is more sensitive to $\taureio$ for our second set of reionization models. This highlights that $\taureio$ depends on both the duration and timing of reionization (through the expansion rate $H(z)$ and the scaling of the proper electron number density with redshift). The sensitivity is also greater to the lower bound on $\taureio$, a feature reflected in the tails of the 1D marginal of $\taureio$ in Figure~\ref{1d}.
However, the lower 1$\sigma$ bound from the \cite{boera19} data depends on the thermal histories considered for the extrapolation in the ${u_0}^{5.0}-{T_0}^{5.0}$ plane, since we do not impose any prior on the former parameter. The lowest ${u_{0}}^{5.0}$ and highest ${T_{0}}^{5.0}$ therefore correspond to the lowest $\taureio$. These \textit{hotter} models with \textit{lower} pressure smoothing are difficult to explain physically.
The 2D contours from the GHOSTLy-like data in Figure~\ref{isocontours}, however, become narrower in the direction constrained by the Lyman-$\alpha$ forest. The lower bound for $\taureio$ is therefore not set by the choice of thermal histories, potentially leading to constraints that are as competitive as those from Planck. 

We have run further an analysis where we impose a prior on $\taureio$ $\geq$ 0.034, which is the lower bound on this parameter assuming instantaneous reionization and $z_{\rm{end}}$ = ${z}_{\rm{end}}^{\rm{P19\,late}}$ (\cite{hu95}). We find $\taureio=0.047^{+0.039}_{-0.012}$ (95\% C.L). The corresponding contours in the ${u_0}^{5.0}-{T_0}^{5.0}$ plane are shown in the \textit{Supplemental Material} \cite{supplemental}. 

The results discussed in this section are subject to a few caveats, mainly introduced by the assumption of Eq.~(\ref{Eion}). If the spectral energy distribution of the sources driving reionization changes with redshift, $E_{\rm{ion}}$ will also be redshift-dependent. This could occur, for instance, in an AGN-assisted reionization model, as has been suggested by recent JWST observations of high-redshift faint AGNs (e.g. \cite{larson23, bogdan24, greene24}). However, these sources have been found to contribute at most $\approx$ 20\% to the reionization budget in order to match Lyman-$\alpha$ forest data \cite{asthana25, dayal25}. Exotic sources of heat injection could further lead to an evolving $E_{\rm{ion}}$($z$), such as dark mater annihilation or decay \cite{liu16}, or cosmic rays remnants of supernovae \cite{savonoz15}.
$E_{\rm{ion}}$($z$) would modify the coupling between $\gammaHI$ and $\epsilon_{\rm{HI}}$, and therefore the correlation between ${u_0}(z)-{T_0}(z)$ and $\taureio$, potentially leading to weaker $\taureio$ constraints. The mapping between these thermal parameters and $\taureio$ relies on an assumed EoR parameterization, which would change if Pop-III stars drive an early on-set of reionization (e.g \cite{wu21,cen03}). We focus on standard reionization histories, as more complex models come with their own caveats.

Finally, we have checked that even a generously estimated   uncertainty of $z_{\rm{end}}$ of $ \sim \pm$0.5 \cite{Gaikwad2023}, shifts the $\taureio$ posterior only by $<$ 1$\sigma$. Marginalizing over this parameter would therefore make little difference to our results.

\textit{Conclusions.---} We have used the Lyman-$\alpha$ forest 1D flux power spectrum and its sensitivity to the thermal state of the IGM, through $u_0$ and $T_0$, to constrain $\taureio$.
Given the coupling between photo-heating and photo-ionization rates, there exists a mapping between the thermal and ionization state of the IGM. We fit a relation between the two by solving the temperature and chemistry evolution equation for hydrogen and helium for two sets of reionization models using the non-equilibrium photoionization code from \cite{bolton22}. We consider $z_{\rm{end}}$-fixed and $\Delta z$-fixed models, the former motivated by Lyman-$\alpha$ forest opacity bounds, the latter inspired by CMB reionization studies. \newline
\indent To find the connection between $u_0-T_0$ and $\taureio$, we use flux power spectrum measurements from \cite{boera19} and find that the redshift bin closest to $z_{\rm{end}}$ is the most sensitive to $\taureio$, as that is when the pressure smoothing effect on the flux power spectrum is largest. Therefore, we fit a relation for $\taureio$ (${u_0}^{5.0}$, ${T_0}^{5.0}$) and use the same analysis framework as in our previous work (\cite{gg25}) to measure $\taureio$. \newline
\indent Our baseline constraint for $z_{\rm{end}}$ and $\Delta z$-fixed, respectively, is ${\taureio} =0.040^{+0.042}_{-0.018}$ and ${\taureio}=0.041^{+0.028}_{-0.017}$. The constraints become stronger when using a mock dataset with the characteristics of the forthcoming GHOSTLy survey. In general, our findings are in agreement with a \textit{rapid} and \textit{late} reionization history from CMB data (\cite{Planck18, cain25, garrett25}) and independent measurements of the ionized hydrogen fraction (\cite{elbersnew}). A large optical depth invoked to solve anomalies in DESI BAO compared to CMB is not supported at the 2.4$\sigma$ and 2.8$\sigma$ for $z_{\rm{end}}$-fixed and redshift-symmetric reionization models, respectively. The statistical significance of this rejection will likely increase with future Lyman-$\alpha$ forest data. \newline
\indent Overall, our findings demonstrate the current constraining power of the Lyman-$\alpha$ forest data on $\taureio$ independently of the CMB, and the level of precision that will be achieved with forthcoming data. Our results, together with observational evidence for a \textit{late} end of reionization from the Lyman-$\alpha$ forest opacity, suggest that $z_{\rm{end}}$-fixed-type of reionization models provide a more physically motivated framework for future reionization studies than redshift-symmetric reionization histories. \\
\indent \textit{Acknowledgments.---}%
The authors thank Laura Keating for useful discussions. 
VI acknowledges partial support by the Kavli Foundation. The simulations used in this work were performed using the Cambridge Service for Data Driven Discovery (CSD3), part of which is operated by the University of Cambridge Research Computing on behalf of the STFC DiRAC HPC Facility (www.dirac.ac.uk).  The DiRAC component of CSD3 was funded by BEIS capital funding via STFC capital grants ST/P002307/1 and ST/R002452/1 and STFC operations grant ST/R00689X/1.  DiRAC is part of the National e-Infrastructure. Support by ERC Advanced Grant 320596 ‘The Emergence of Structure During the Epoch of Reionization’ is gratefully acknowledged. OG has been supported by STFC grant  ST/Y509139/1. MGH has been supported by STFC consolidated grants ST/N000927/1 and ST/S000623/1. JSB is supported by STFC consolidated grant ST/X000982/1.


\bibliography{references}

\onecolumngrid

\end{document}